# Phase-evolution in Co-Sb System: CoSb$_3$ Hydrothermal Synthesis

Kanika Upadhyay[a], Sanjeev Gautam[b,*] and Navdeep Goyal[c,*]

[a]*Department of Physics, Panjab University, Chandigarh 160014, India*.
[b]*Dr. SSB University Institute of Chemical Engg. & Tech, Panjab University, Chandigarh - 160014, India*

**Abstract**

Nanostructure CoSb$_3$ samples were synthesized via hydrothermal technique along with different routes to study the phase-evolution in Co-Sb system. X-ray diffraction (XRD) explored to study the phase evolution in Co-Sb system via. Special emphasis on rietveld refinement and scanning electron microscope (SEM) were evaluated to study morphology of the samples. The samples prepared using ramp-route showed presence of different phases that eventually helps in formation of pure phase CoSb$_3$ skuterrudite. Using different solvents to eliminate NaCl impurity also shows that DI-water eliminates NaCl impurity better because of its high solubility.

## 1. Introduction

Research on Thermoelectric (TE) materials has taken its pace in the past few decades owing to global energy crisis and degradation of fossil fuels [1]. Various materials have been studied in quest of potential TE materials, Skutterudites (MX$_3$, M= Tm; X= As, Sb) being capable of showing merits in mid-temperature power generation is one of the most studied materials [2, 3]. The CoSb$_3$ skutterudite due to its reasonable band gap, high carrier mobility and thermal stability have proven to be a strong candidate among its family and other TE materials. The high thermal conductivity of bare CoSb$_3$ limits its commercial applicability [4, 5]. The performance of TE device depends upon material efficiency and device design [6]. Furthermore, the efficiency of TE material depends upon figure of merit, given by:

$$ZT = S^2\sigma T/k \qquad (1)$$

Where S is Seebeck coefficient, σ is electrical conductivity, T is temperature and κ is thermal conductivity. For a better TE material one need to increase the power factor along with decrease in its thermal conductivity and as the properties of a material directly depends on the synthesis technique adopted. This calls a need to search for a specific nanoparticle synthesis route.

The Co-Sb system exists in three different phase i.e CoSb, CoSb$_2$ and CoSb$_3$ and for TE applicability synthesis of CoSb3 in pure form is required as presence of mono and diantimonide can deteriorate the performance of the material and filling fraction limit of doped skutterudite [7]. Several attempts have been made for the synthesis of pure phase CoSb$_3$. Alinejad *et al.* synthesized single phase CoSb$_3$ with peritectic solidification concept i.e. heating the Co and Sb powder in encapsulated tube firstly at 1200 °C for monoantimonide formation and then at 940 °C and 885 °C that are the two peritectic points of Co-Sb system for diantimonide and finally pure phase triantimonide formation [8]. In contrast, Mi *et al.* and Gharleghi *et al.* attempted to synthesize pure phase CoSb$_3$ using solvothermal method, as it optimizes the ambient conditions and is more controllable as compare to other conventional methods. They also synthesized the samples at different temperature for different synthesis duration and suggested that formation of diantimonide thermodynamically dominates the formation of triantimonide [9] and also observed the coexistence of all phases of Co-Sb system along with Sb as impurity [10]. To get rid of the unwanted impure phases, it is of utmost importance to know the growth mechanism of Co-Sb system, so that a pure and economic solution is provided to thermoelectric materials.

In this research, a systematic hydrothermal synthesis process is performed to study the different phase evolution in Co-Sb system. Three different samples of CoSb$_3$ were prepared by heating the autoclave directly in which initial chemicals were mixed at 210 °C for 90 hours and in ramp-manner for 40 and 60 hours respectively. Sample prepared for 60 hours was further washed with DI-water for removal of NaCl impurity which was earlier washed with ethanol only.

## 2. Experimental details

Analytical grade $CoCl_2 \cdot 6H_2O$ and $SbCl_3$ were weighed and added in 15 ml ethanol to prepare 0.1 and 0.3 molar solutions respectively. The solutions were then loaded in the teflon container and sonicated for 10 minutes for proper mixing. Sufficient amount of $NaBH_4$ was added to the above solution in dropwise manner for proper reduction. The container was placed in the autoclave and sealed properly which was then subjected to temperature 210 ºC in the oven directly for 90 hours and in ramp-route for 40 and 60 hours respectively and autoclave was then brought to room temperature naturally. Figure 1 represents the time-temperature curve followed for ramp-route synthesis. The obtained products were filtered out and washed several times with ethanol and 60 hours sample was also washed with de-ionized (DI) water. The samples were then dried at 100 ºC for 12 hours. Samples so prepared for 90 hours, 40 hours, 60 hours and DI water washed were labeled as CS-90, CS-40, CS-60 and CS-60D respectively.

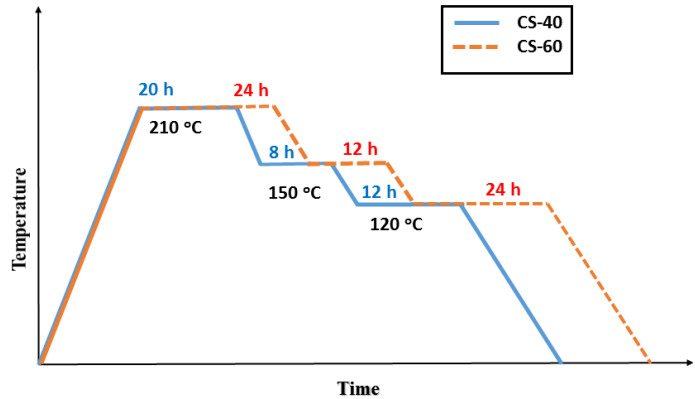

Figure 1 : Time-temperature curve for CS-40 and CS-60 ramp-route synthesis.

To study the phase evolution in Co-Sb system the samples were analyzed by Rigaku X-ray diffractometer equipped with Cu-K$_α$ radiation (1.5404 Å) from 10º-80º range with rotation speed 1º/min to get maximum intensity. The Rietveld refinement studies of the sample XRD profiles were performed using X'pert HighScore Plus. The morphology and elemental impurity of the samples were examined using Scanning Transmission Electron Microscope (SEM), Model JSM6100 (Jeol) with Image Analyser.

## 3. Results and discussions

Figure 2 shows the XRD pattern of the powder samples prepared through hydrothermal technique. No obvious peaks of $CoSb_2$ and $CoSb_3$ phase were observed in CS-90 pattern, whereas in CS-40 and CS-60 pattern, peaks corresponding to $CoSb_2$ and $CoSb_3$ were detected. The observation indicates that heating the autoclave directly at 210 ºC for 90 hours

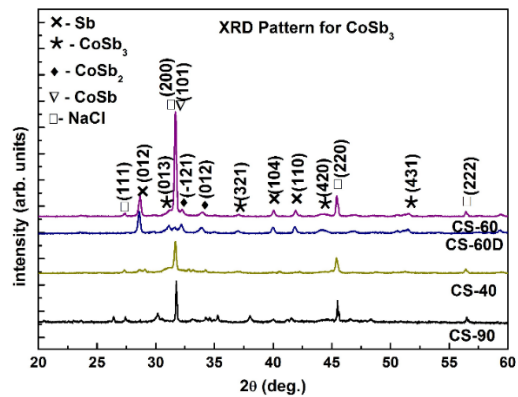

Figure 2: Powdered XRD pattern of hydrothermally synthesized $CoSb_3$ with different synthesis profiles.

led only to the production of mono-antimonides, while the ramp-route provide sufficient amount of time for the production of CoSb, $CoSb_2$, $CoSb_3$ phase. The peaks of CoSb, $CoSb_2$, $CoSb_3$, Sb and NaCl were indexed using JCPDS

number [00-033-0097], [01-089-2845], [01-089-4871], [01-085-1322] and [01-070-2509] respectively. From Fig.2 it was also observed that the samples washed with ethanol still had NaCl (at 27.34, 45.43 and 56.47) as an impurity whereas DI- washed sample showed no trace of NaCl. Using Scherrer's equation the crystallite size of the CS-60 prepared sample was calculated to be 37.13 nm.

The Reitveld refinement of CS-90 and CS-60D was performed to analyze the amount of each phase present which will further help in comprehending phase transformation in different synthesis profiles. Fig.3 (a) and (b) represents the Reitveld refinement of CS-90 and CS-60D respectively along with table 1 of R-factors and percentage of each phase present obtained from each refinement. In X'pert high score software firstly the background for each sample was determined manually followed by peak search and loading of reference data of expected phase present. The data was then converted to phase and after that detail of .cif data were loaded in the atomic coordinates of each phase in refinement control panel. Refinement was then performed in step wise manner. The background was refined using polynomial and the peak shape was refined through Pseudo-Voigt function. It can be seen from the table of R-factor that the value of $R_{profile}/(\%)$, $R_{weighted\ profile}/(\%)$ and $R_{expected}/(\%)$ were less than 10% and GOF less than 4% which justifies a very good refinement. All the percentage of each phase along with impurity present has also been represented with the help of a pie-chart.

The pie chart shows NaCl as the major phase in CS-90 on other hand no such trace has been observed in CS-60D. This shows that washing the sample with DI-water eliminates the presence of NaCl better than the one where the samples were washed with ethanol instead. It was also observed form table 1 that percentage of mono and diantimonide increased from 10.5% to 12.3% and 11.7% to 23.4% respectively. Cobalt triantimonide phase with percentage of 35.9% was also detected. The peaks marked in Fig. 3 (a) are peaks of unknown phases which was not refined during Reitveld refinement. The presence of $B_5H_9$ in CS-90 might have arisen as Teflon container was immediately sealed after the reduction process without allowing $BH_3$ to escape. $BH_3$ might have further reacted in the autoclave at high temperature and pressure and led to the production of $B_5H_9$

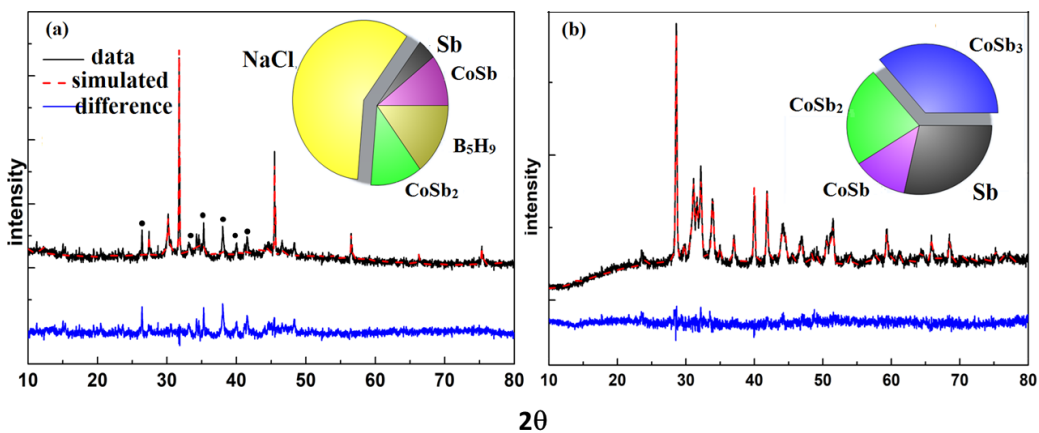

Figure 3: Rietveld refined XRD plot of (a) CS-90 and (b) CS-60D

| Phase/Sample | | CS-90 | CS-60D |
|---|---|---|---|
| Percentage | $CoSb_3$ | ----- | 35.9 |
| | $CoSb_2$ | 11.7 | 23.4 |
| | CoSb | 10.5 | 12.3 |
| | Sb | 4.2 | 28.4 |
| | NaCl | 58.9 | ----- |
| | $B_5H_9$ | 14.6 | ----- |
| $R_{profile}$ | | 3.13 | 5.84 |
| $R_{weight-profile}$ | | 4,25 | 7.41 |
| $R_{expected}$ | | 2.89 | 5.84 |
| GOF (goodness of fit) | | 2.17 | 1.61 |

Table 1: R-factor and percentage of each phase present in CS-90 and CS-60D

The exact mechanism for formation of CoSb$_3$ is not clear yet as per the literature available. Using XRD analysis, we can say that the formation process for CoSb$_3$ undergoes multiple steps:

$$CoCl_2 + 2NaBH_4 \rightarrow Co + 2BH_3 + 2NaCl + H_2 \quad (2)$$

$$2SbCl_3 + 6NaBH_4 \rightarrow 2Sb + 6BH_3 + 6NaCl + 3H_2 \quad (3)$$

$$Co + Sb \rightarrow CoSb \quad (4)$$

$$CoSb + Sb \rightarrow CoSb_2 \quad (5)$$

$$CoSb_2 + Sb \rightarrow CoSb_3 \quad (6)$$

The co-existence of CoSb, CoSb$_2$ and CoSb$_3$ in CS-40 and CS-60D suggest that CoSb$_3$ skutterudite is formed from the addition of antimony into monoantimonide and then diantimonide molecule rather than direct combination of one Co atom and three Sb atoms. The persistent presence of Sb phase in the samples indicates sufficient temperature is required for the antimony atom to react with mono and diantimonide to form triantimonide along with ramp-route synthesis. The conclusion that CoSb and CoSb$_2$ acted as intermediate product for formation of CoSb$_3$ matches with the work of Xie *et al.* [11].

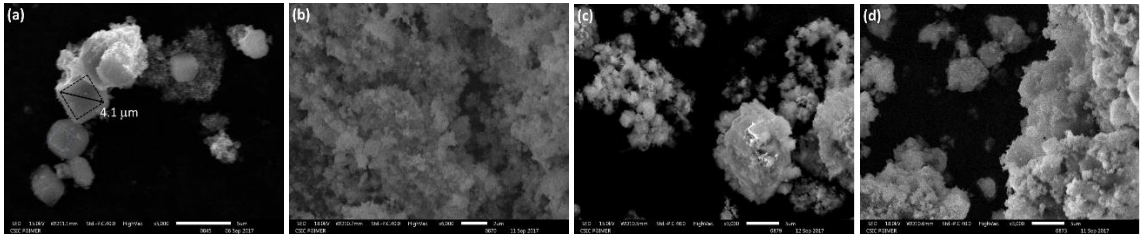

Figure 4: SEM micrographs of (a) CS-60, (b) CS-60D, (c) CS-90 and (d) CS-60D

Figure 4 presents the SEM micrographs of the CoSb$_3$ nanoparticles prepared using different routes. The CS-60 synthesized sample has grain size of 1.82 µm, where most of the granules form irregular shape, as CoSb$_3$ and NaCl both possess cubic structure, it is difficult to differentiate between these two structures. With increase in synthesis duration, the morphology of the synthesized nanostructure has not been changed. However, from Fig. 4(a) and (b), it can be concluded that after the elimination of NaCl impurity agglomeration in the sample occurs. Fig 4(c) shows no traces of CoSb$_3$ phase, which agrees with the XRD data obtained.

**4. Conclusions**

In summary, different routes followed for the preparation of CoSb$_3$ nanoparticle via hydrothermal method helps in understanding the phase-evolution in Co-Sb system. The structural analysis using Rietveld refinement confirms that the formation of CoSb$_3$ phase via direct route is difficult although using ramp-route formation of cubic phase of CoSb$_3$ can be achieved. The highest percentage of CoSb$_3$ phase was observed in case of CS-60D with 35.9%. Using hydrothermal method sample with 37.1 nm crystallite size is attained. XRD analysis also shows that the formation of monoantimonides and diantimonides acts an intermediate product for formation of CoSb$_3$ skutterudite and synthesis of pure phase CoSb$_3$ can be attained by increasing the synthesis temperature > 240 ºC along with ramp-route method.

**Acknowledgements**

Kanika acknowledges the CAS-JRF fellowship from UGC. The research is partially supported by UGC-BSR and TEQIP- III R&D research project.

# References

[1] H. Alam and S. Ramakrishna, Nano Energy 2(2013) 190-212.
[2] T. Caillat, A. Borshchevsky, and J.-P. Fleurial, J. Appl. Phys. 80(1996) 4442–4449.
[3] J. O. Sofo and G. D. Mahan, Phys. Rev. B 58(1998) 15620–15623.
[4] L. Tayebi, Z. Zamanipour, and D. Vashaee, Renewable Energy 69(2014) 166 – 173.
[5] H. Ehrenreich and F. Spaepen, eds., Solid State Physics, Vol. 51 (Academic Press, 1997).
[6] Lobat Tayebi, Zahra Zamanipour, Daryoosh Vashaee, Renewable Energy, 69 (2014) 166-173
[7] X. Shi, W. Zhang, L. D. Chen, and J. Yang, Phys. Rev. Lett., 95(2005) 185503
[8] Babak Alinejad, Alberto Castellero, Marcello Baricco, Scripta Materialia, 113(2016) 110-113
[9] J.L. Mi, X.B. Zhao, T.J. Zhu, J.P. Tu, G.S. Cao, J. Alloys Compds. 417 (2006) 269–272
[10] Ahmad Gharleghi, Yi-Hsuan Pai, Fei-Hung Lin and Chia-Jyi Liu, J. Mater. Chem. C, 2 (2014) 4213–4220
[11] J. Xie, X.-b. Zhao, J.-l. Mi, G.-s. Cao, and J.-p. Tu, Journal of Zhejiang University-SCIENCE A 5(2004) 1504–1508.